%% file: main.tex
\RequirePackage{lineno}
\documentclass[aps,prdrc,twocolumn,superscriptaddress,groupedaddress,floatfix]{revtex4}  

\usepackage{graphicx}  
\usepackage{dcolumn}   
\usepackage{comment}
\usepackage{bm}        
\usepackage{amssymb}   
\usepackage{amsmath}
\usepackage{booktabs} 
\usepackage{braket}
\usepackage{breqn}
\usepackage[normalem]{ulem}
\usepackage{siunitx}

\usepackage{xcolor}
\usepackage{xspace, tabularx}
\usepackage{caption}
\usepackage{threeparttable}

\newcommand{\beq}{\begin{equation}}
\newcommand{\eeq}{\end{equation}}
\newcommand{\beqa}{\begin{eqnarray}}
\newcommand{\eeqa}{\end{eqnarray}}

\newcommand\barbracket[1]{\overset{%
   \scriptscriptstyle(-)}{#1}}
\newcommand{\numuParen}{\ensuremath{{\barbracket{\nu}_{\mu}}}\xspace}

\begin{document}

\title{Search for active-sterile antineutrino mixing using neutral-current interactions with the NOvA experiment}

\input novancsterile20.tex


\begin{abstract}
This Letter reports results from the first long-baseline search for sterile antineutrinos mixing in an accelerator-based antineutrino-dominated beam.  The rate of neutral-current interactions in the two NOvA detectors, at distances of 1\,km and 810\,km from the beam source, is analyzed using an exposure of $12.51\times10^{20}$\,protons-on-target from the NuMI beam at Fermilab running in antineutrino mode.  A total of 121 of neutral-current candidates are observed at the Far Detector, compared to a prediction of $122\pm11\textnormal{(stat.)}\pm15\textnormal{(syst.)}$ assuming mixing only between three active flavors.  No evidence for $\bar{\nu}_{\mu}\rightarrow \bar{\nu}_s$ oscillation is observed.  Interpreting this result within a 3+1 model, constraints are placed on the mixing angles $\theta_{24} < 25^{\circ}$ and $\theta_{34} < 32^{\circ}$ at the 90\% C.L. for $0.05\,\textnormal{eV}^2 \le \Delta m_{41}^2 \le 0.5\,\textnormal{eV}^2$, the range of mass splittings that produces no significant oscillations at the Near Detector.  These are the first 3+1 confidence limits set using long-baseline accelerator antineutrinos.
\end{abstract}

\maketitle


Studies of neutrinos and antineutrinos from a variety of sources, including accelerators, the atmosphere, the Sun and nuclear reactors~\cite{ref:SuperK1998,ref:SNO2002,ref:KamLAND2005,ref:K2K2006,ref:MINOS2006,ref:T2K2011,ref:DayaBay2012,ref:RENO2012,ref:DoubleChooz2014,ref:OPERA2015,ref:NOvA2016_PRL}, provide substantial evidence for mixing between the three known active flavors $\barbracket{\nu}_e$, \numuParen and $\barbracket{\nu}_{\tau}$.  However, the observation of antineutrinos in short-baseline oscillations~\cite{ref:LSND2001,ref:MiniBooNE2018}, reactor~\cite{ref:Huber2011,ref:Mueller2011} and gallium-based radiochemical experiments~\cite{ref:Acero-Ortega,ref:Giunti} found inconsistency with this 3-flavor model.  These anomalous results may be explained by extending the mixing model to include more massive neutrino and antineutrino states in addition to $\barbracket{\nu}_1$, $\barbracket{\nu}_2$ and $\barbracket{\nu}_3$ with new mass splittings large enough to provide oscillations over much shorter baselines than is possible with the known 3-flavor mass differences.  However, observations of the width of the $Z^0$-boson resonance at the LEP experiments put the constraint on the number of light antineutrinos participating in the weak interaction to be three~\cite{ref:pdg}.  Any additional antineutrinos of mass less than approximately half the $Z^0$ mass must therefore be sterile~\cite{ref:SterileWhitePaper}.

The simplest extension to this 3-flavor model is referred to as the 3+1 model and introduces a single new mass state, $\barbracket{\nu}_4$, with a corresponding sterile flavor state, $\barbracket{\nu}_s$, such that $\ket{\barbracket{\nu}_s} = \sum_{i=1}^4 U_{si}^*\ket{\barbracket{\nu}_i}$,
where $U$ represents a unitary $4\times4$ extended Pontecorvo-Maki-Nagakawa-Sakata (PMNS) matrix~\cite{ref:Pontecorvo1958,ref:MNS1962}.  The matrix can be parametrized as $U = R_{34}S_{24}S_{14}R_{23}S_{13}R_{12}$ \cite{ref:matrixpram}, where $R_{ij}$ represents a rotation through mixing angle $\theta_{ij}$ and $S_{ij}$ represents a complex rotation through angle $\theta_{ij}$ and phase $\delta_{ij}$.  In addition to the 3-flavor mixing parameters and a new independent mass splitting $\Delta m_{41}^2 = m_4^2-m_1^2$, this model includes three new mixing angles, $\theta_{14}$, $\theta_{24}$ and $\theta_{34}$, and two additional phases, $\delta_{14}$ and $\delta_{24}$, which may violate CP symmetry.  A number of atmospheric, long-baseline accelerator and reactor neutrino experiments have searched for oscillations outside of the 3-flavor mixing framework and found no evidence of such processes~\cite{ref:MINOS2019,ref:DayaBaySterile2015,ref:SuperKSterile2015,ref:IceCube2016,ref:DeepCore2017,ref:T2KSterile2019,ref:NOvASterile2017}, placing constraints on the allowed values of the parameters governing 3+1 oscillations.  Equivalent constraints have not yet been published from similar investigations of antineutrinos, the focus of the analysis described in this Letter.  This is pertinent as the discrepancies observed in LSND~\cite{ref:LSND2001} and MiniBooNE~\cite{ref:MiniBooNE2010} involved measurements of oscillations from antineutrino beams, with different behavior reported in subsequent analyses of oscillations of neutrinos~\cite{ref:MiniBooNE2013}.  Further investigation is required to understand the oscillations of antineutrinos into potential sterile states.

NOvA can search for evidence of active to sterile oscillations through an analysis of the rate of neutral-current (NC) interactions in its Near Detector (ND) and Far Detector (FD)~\cite{ref:NOvASterile2017}.  When oscillations occur between only the three active flavors, the rate of NC interactions is independent of flavor and is thus unaffected by oscillations.  If the antineutrinos transition into an unseen sterile flavor state during propagation, a reduction in the number of NC interactions would be observed with a probability given approximately by~\cite{ref:MINOS2016}

\begin{align}\label{eq:OscProb}
\begin{split}
    1 - P\left(\barbracket{\nu}_{\mu}\rightarrow\barbracket{\nu}_s\right)\approx 1 &- c_{14}^4c_{34}^2\sin^2{2\theta_{24}}\sin^2{\Delta_{41}} \\
    &- A\sin^2{\Delta_{31}} \pm B\sin{2\Delta_{31}},
\end{split}
\end{align}
where $c_{ij}=\cos{\theta_{ij}}$, $\Delta_{ji}=\frac{\Delta m^2_{ji}L}{4E}$ and $\Delta m^2_{ji}=m^2_j-m^2_i$.
To first order, $A=\sin^2{\theta_{34}}\sin^2{2\theta_{23}}$ and $B=\frac{1}{2}\sin{\delta_{24}}\sin{\theta_{24}}\sin{2\theta_{34}}\sin{2\theta_{23}}$.  This approximation assumes $\Delta_{21}=0$, $\sin^2{2 \theta_{23}}$ is near-maximal, and considers the fast oscillation regime with limited detector resolution such that $\sin^2{\Delta_{41}}=\frac{1}{2}$; an exact formalism, including matter effects, is used in the analysis.  Studying the disappearance of NC events over a long baseline is therefore sensitive to the 3+1 parameters $\theta_{24}$, $\theta_{34}$, $\delta_{24}$ and $\Delta m^2_{41}$.

Differences between neutrino and antineutrino oscillations due to CP-violation in this approximation are described by the $\delta_{24}$ phase, and therefore the sign of the $B$ term, and are at most around 10\% in the peak antineutrino flux region (around 2\,GeV), with a maximum of 20\% at higher energies.  While an additional asymmetry results from neutrinos gaining an extra effective mass due to forward scattering in matter, this effect is significantly smaller than from the phase.

The $\Delta m^2_{41}$ parameter determines the frequency of the oscillations.  For sufficiently small values, the oscillations develop over longer baselines than the ND and are more rapid than can be resolved at the FD, resulting in an average reduction in the NC rate over the energy spectrum.  As the mass splitting increases, oscillations develop over shorter baselines and are evident at the ND.  This analysis considers oscillations driven by $0.05\,\textnormal{eV}^2 \le \Delta m_{41}^2 \le 0.5\,\textnormal{eV}^2$, where the NC rate is unaffected at the ND and reduced at the FD.

The NOvA experiment consists of two functionally identical detectors placed in the NuMI beam~\cite{ref:NuMI2016}.  The ND is located at Fermilab in Batavia, Illinois, 1\,km from the NuMI target and 100\,m underground; the FD is located near Ash River, Minnesota, 810\,km from the beam source with a 3\,m water-equivalent overburden of concrete and barite.  The detectors are placed 14.6\,mrad off the axis of the beam to produce a narrow neutrino or antineutrino energy distribution peaked at 2.0\,GeV at the FD with a width of 0.4\,GeV and a high energy tail of around 10\% of the peak.

The NuMI beam is produced from 120\,GeV protons incident on a mostly carbon target.  Two magnetic focusing horns pulsed at 200\,kA focus the secondary pions and kaons, which subsequently decay in flight over 705\,m, including a 675\,m decay pipe, to mainly muons and muon (anti)neutrinos.  By selecting the current polarity in the horns, the beam can be run in a mode dominated by either $\nu_\mu$ or $\bar{\nu}_{\mu}$.  The beam is extracted for 10\,$\mu$s approximately every 1.3\,seconds.  A previous analysis searched for evidence of NC-disappearance in a $\nu$-dominated beam~\cite{ref:NOvASterile2017}; this study is the first using the NOvA $\bar{\nu}$-dominated dataset.  Across the 0-20\,GeV energy range, beam interactions at the ND are predicted to consist of 63\%~$\bar{\nu}_\mu$, 35\%~$\nu_{\mu}$ and 2\%~$\nu_e+\bar{\nu}_e$.  In the 0-5\,GeV peak energy range, the region which provides greatest sensitivity to $\theta_{34}$ in this study, beam interactions are 81\%~$\bar{\nu}_\mu$.  The analysis signal is NC events, with the sample dominated by $\bar{\nu}$ but with some $\nu$ component that is treated the same; all charged-current (CC) events are considered background.

The detectors are constructed of highly reflective PVC cells~\cite{ref:PVC} with cross-section 3.9\,cm-by-6.6\,cm and 15.5~(3.8)\,m in length for the FD~(ND). The cells are formed into 896~(214) planes in the FD~(ND), alternating horizontal and vertical relative to the beam axis to enable three-dimensional tracking.  Each cell is filled with mineral-oil based liquid scintillator doped with 5\% pseudocumene~\cite{ref:Oil} and instrumented with a loop of wavelength-shifting fiber, read out at one end using a 32\,pixel avalanche photodiode~\cite{ref:apd}.  The FD and ND have masses of 14\,kt and 193\,t, respectively.  Custom readout electronics shape and digitize the signal.  Pulse height and timing of all energy deposits above a given noise threshold are collected in a 550\,$\mu$s window encompassing the NuMI beam spill at both detectors.  Data are additionally collected at the FD using a 10\,Hz minimum bias trigger to sample the cosmogenic background.

The analysis uses data recorded between June 2016 and February 2019, corresponding to an exposure of $12.51\times10^{20}$ protons-on-target (POT) for the FD.  The amount of ND data used corresponds to an exposure of $3.10\times10^{20}$ POT.

A full simulation of the beam, including all materials and interactions in the beamline, is used to predict the flux at the NOvA detectors.  Simulation of the antineutrino production from primary proton-induced hadronic showers and subsequent meson transport and decay is provided by a custom made simulation based on GEANT4~\cite{ref:GEANT2003,ref:GEANT2006} with a full geometry of the beamline.  The flux is then corrected using a version of the Package to Predict the Flux (PPFX)~\cite{ref:AliagaThesis} for the NOvA off-axis beam~\cite{ref:NOvAFlux}, which modifies the particle production using external hadron production data. 

Neutrino interactions in the NOvA detectors are simulated by GENIE~\cite{ref:GENIE2010}, with modifications performed by NOvA~\cite{ref:NOvAXSecTune2020}.  The particles are propagated through the detector geometry using GEANT4; the charge deposits are converted to scintillation and transported, along with Cherenkov light, to the front-end electronics using a custom made simulation process.  The response of the photo-detectors is well understood and used to simulate charge collected by the front-end electronics.

NC events are characterized by the lack of a charged lepton exiting the vertex and typically consist of diffuse hadronic activity resulting from energy and momentum transfer from the antineutrino to the nucleus.  Interactions are identified by requiring a level of activity consistent with these topologies and rejecting CC events with obvious antimuon tracks or positron showers.  The NOvA detectors are well-suited to identify NC interactions since the fine granularity results in the hadronic interaction length of 38\,cm extending to around 10\, cells.  This enables a good distinction between $\pi^0$s and positrons, important in selecting NC events containing hadronic activity and rejecting $\bar{\nu}_e$~CC interactions~\cite{ref:NOvATDR}.

The selection was developed on simulated interactions and tested on the ND data, with identical requirements applied to each.  The FD has far less overburden than the ND and is exposed to a much higher rate of cosmic-induced interactions; supplementary criteria are therefore applied to FD data to remove the cosmogenic backgrounds.  The selection applied in this study, outlined below, is based on the previous neutrino analysis described in~\cite{ref:NOvASterile2017}.

Hits, cells with activity above threshold, are grouped based on their proximity in space and time to form neutrino events candidates.  An event vertex is then reconstructed from these hit collections by projecting back to a common start point.  Events are required to have a vertex and be fully contained in the detector.  This containment requirement stipulates no activity within an outer volume defined for the ND as 25\,cm from all sides and for the FD as 100\,cm from the top of the detector, 10\,cm from the bottom and 50\,cm from each of the other sides.  Backgrounds arising from neutrino interactions in the material surrounding the ND and resulting in particles, such as neutrons, entering and interacting in the detector are removed by requiring the reconstructed vertex be greater than 100\,cm from the side walls and between 150\,cm and 1000\,cm from the upstream face of the detector.

\begin{figure}
    \centering
    \includegraphics[width=\linewidth]{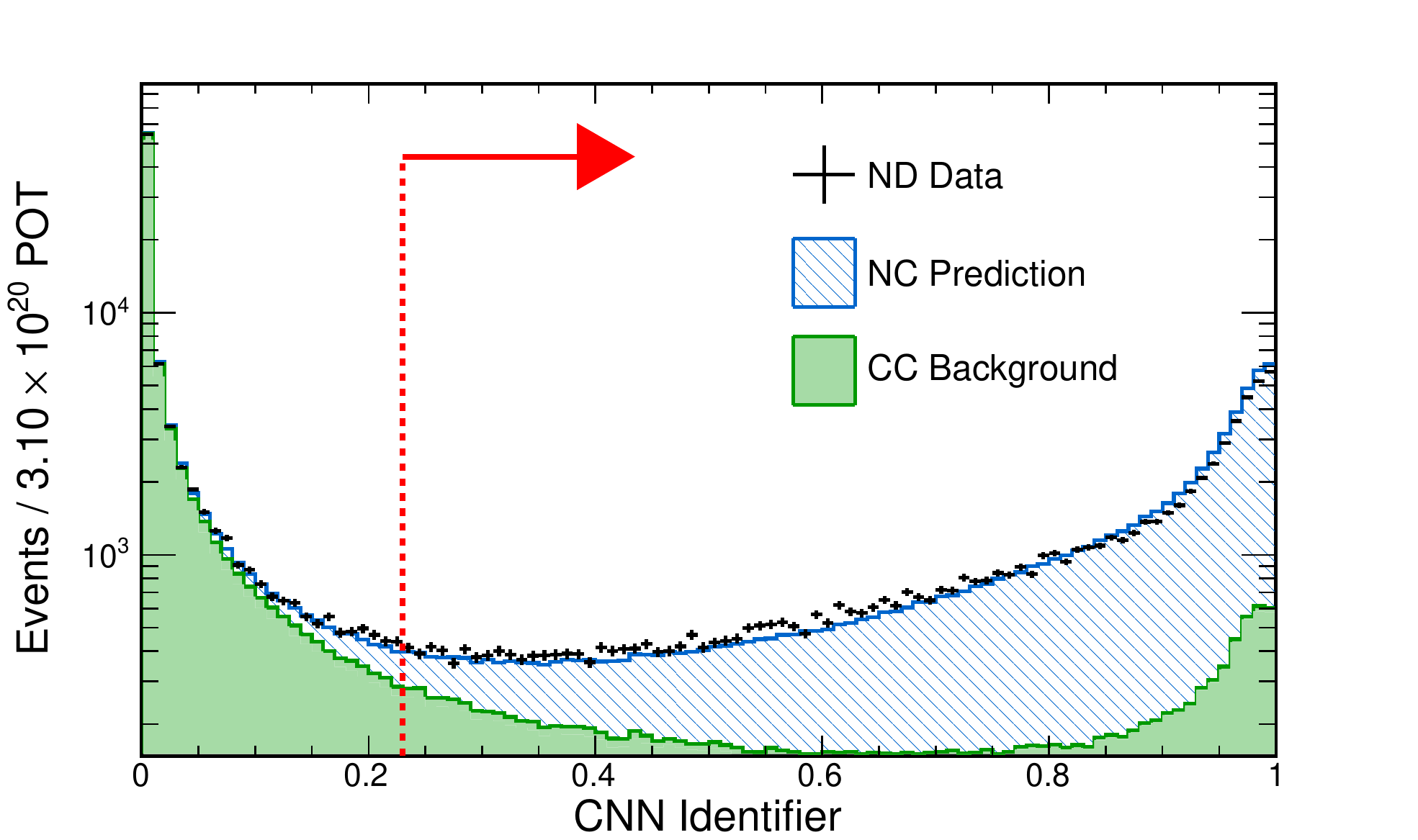}
    \caption{The CNN likelihood output for neutral-current events used to distinguish between various interaction topologies applied to ND simulation and data.  The signal and the dominant predicted background from CC beam interactions are shown.  The red arrow indicates the selected events.}
    \label{fig:CVN}
\end{figure}

The primary event classification is provided by a NOvA-custom Convolutional Neural Network (CNN)~\cite{ref:CVN2016} which identifies interaction topologies from images of the two orthogonal detector views.  It is trained separately on simulated FD neutrino- and antineutrino-dominated beam events and cosmic-ray data, and distinguishes the final-state topologies between $\barbracket{\nu}_{\mu}$~CC, $\barbracket{\nu}_e$~CC, $\barbracket{\nu}_{\tau}$~CC, NC, and cosmic-induced events.  The network learns to extract the characteristic features of the events described above, for example, identifying interactions with charged leptons exiting the vertex as CC, and does not distinguish between interactions from neutrinos and antineutrinos.  The separation provided by the CNN for NC events is shown in Fig.~\ref{fig:CVN}, and a cut at 0.23 is applied to select events in both detectors.  The peak of background events close to 1 is comprised primarily of high energy deep inelastic scattering interactions, or low energy events with minimal detector activity, neither with an obvious lepton to identify.  Interactions occurring in the surrounding rock and subsequently entering the ND are removed by requiring the fractional component of the event momentum which is transverse to the beam direction to be less than 80\%.  At the FD, similar criteria requiring this to be less than 40\% for showers with a vertex within 2.4\,m from the top of the detector and less than 20\% for a vertex inside 2\,m are used to remove cosmic-induced interactions.  For the FD, candidate antineutrino events must not have significant additional activity within 5\,m and 50\,ns of the final state particles.  Events with activity within 200\,cm from the end of the detector must have a sufficient number of downstream planes with hits, compared to the front planes, to ensure showering particles are in the forward direction.  A boosted decision tree that considers multiple shower properties is used to further remove cosmic-induced interactions at the FD~\cite{ref:YangThesis}.  Following the application of the full selection, the cosmic background is reduced by over six orders of magnitude.  The selected sample in the FD simulation is 78\% pure.  Two-thirds of the remaining backgrounds result from CC interactions where most of the energy is transferred to the hadronic system, and the rest originate from cosmic-induced interactions.  The NC-candidate events selected from beam neutrino interactions are 79\%~$\bar{\nu}$ and 21\%~$\nu$ in the 0-5~GeV peak flux region.

The antineutrino-candidate energy is estimated for NC events by summing the calibrated energy deposited in the interaction and applying a linear scaling from simulation to correct for the unseen outgoing energy.  The final reconstructed energy has a resolution of around 30\%.

\begin{figure}[ht]
    \centering
    \includegraphics[width=\linewidth]{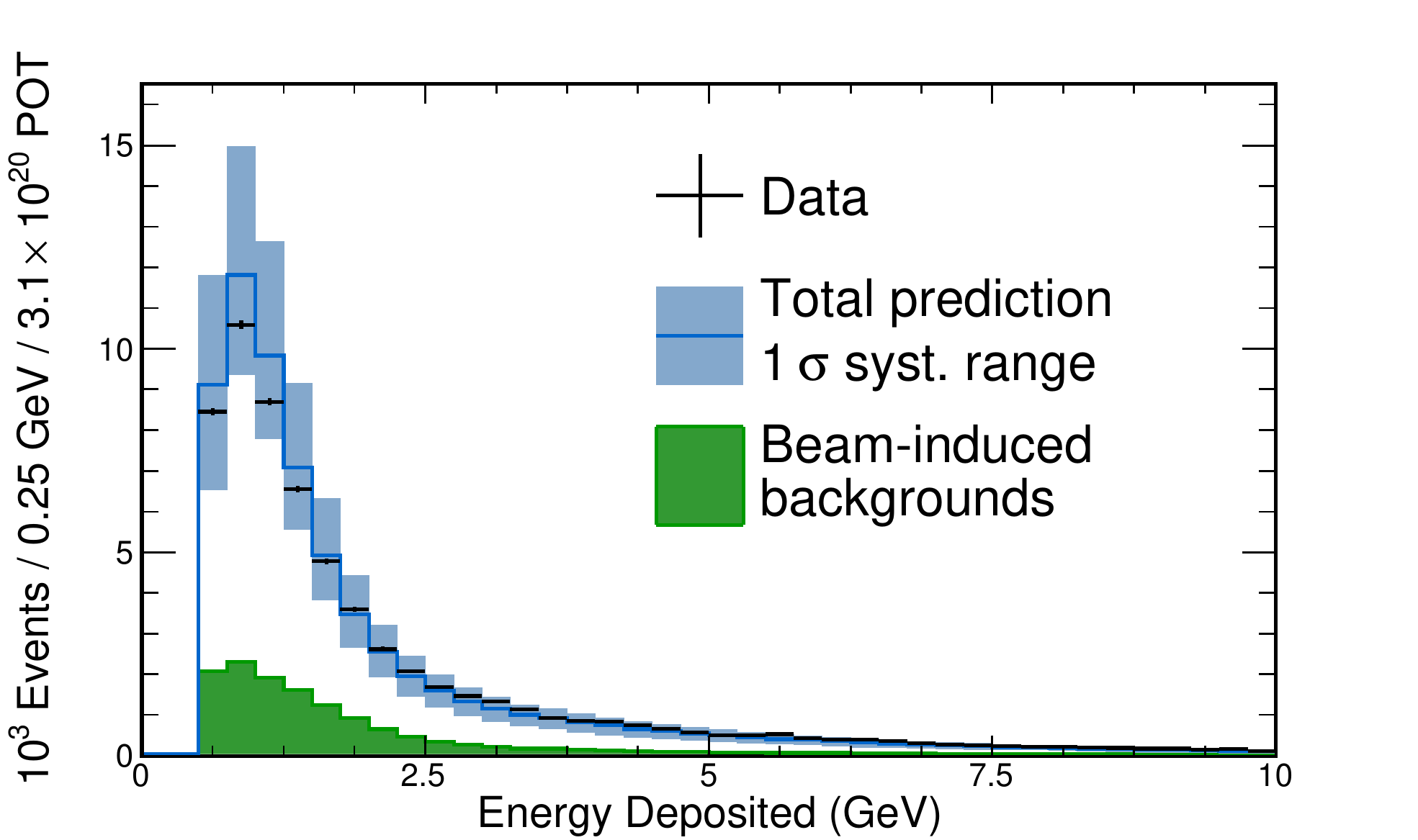}
    \caption{Spectrum of ND events passing the NC selection.  The shaded band shows the total systematic uncertainty.  The solid-fill component histogram displays the background contribution from beam CC interactions.}
    \label{fig:NDSpectrum}
\end{figure}

The spectrum of selected NC-candidate events at the ND is shown in Fig.~\ref{fig:NDSpectrum}.  From this, the predicted spectrum at the FD is produced following a data-driven approach in which the ND data are used to constrain the selected simulated events before propagating the resulting distribution to the FD~\cite{ref:NOvA2018}.  This procedure uses the simulated ratio between the spectra at the two detectors and takes into account geometric differences and effects of the beam dispersion.  The flavor composition and energy of the corrected ND NC-candidate spectrum is inferred from simulation and used to apply oscillations to the events.
This approach is effective in partially cancelling any correlated uncertainties between the two detectors but results in sensitivity only to active-to-sterile oscillations which develop over larger distances than the ND baseline.  In a 3+1 model these slower oscillations are driven by mass splittings in the range $0.05\,\textnormal{eV}^2 \le \Delta m_{41}^2 \le 0.5\,\textnormal{eV}^2$.

The dominant uncertainties are summarized in Table~\ref{tab:SystSummary}.  The largest are attributed to the calibration process which determines the energy scale of the charge collected in each detector cell.  An absolute calibration uncertainty of 5\% is used, consistent with the agreement with simulation of proton response in quasielastic-like ND CC interactions.  A calibration shape uncertainty to cover charge deposited near the poorly-modeled detector edges is evaluated by considering the differences between calibrated and simulated deposits.  These uncertainties are applied both as correlated absolute offsets at both detectors and separately as relative differences between them, since they are determined using just the ND.  Reducing this large calibration uncertainty is a priority for subsequent analyses and there are various ongoing efforts, such as improving the analysis procedure and utilizing understanding gained from a test beam program \cite{ref:WallbankTB2020}.

The simulation contains a modeling of the detector response, including the creation of scintillation and Cherenkov light, its transport to the front-end electronics and subsequent processing.  An uncertainty of 10\% in the normalization of the scintillation light is determined from differences between data and simulation in the number of photoelectrons collected by through-going minimally-ionizing particles.  The residual uncertainties are quantified using alternate predictions where the scintillation and Cherenkov photon yields in the model are altered within the tolerance of agreement with the ND data while holding the muon response fixed, since that is set by the calibration procedure.

Neutrino interaction model uncertainties are estimated using the GENIE reweighting framework, with additional uncertainties on the modifications of the various interaction models performed by NOvA~\cite{ref:NOvAXSecTune2020}.  An additional uncertainty is considered to account for the differences between the selected spectrum with and without applying these NOvA reweighting modifications.  A further 60\% uncertainty on the $\nu_{\tau}$~CC cross-section is allowed, consistent with the most recent measurements~\cite{ref:DONuT2008,ref:SuperKTau2018,ref:OPERA2018,ref:IceCubeTau2019}.

Uncertainties in particle propagation when simulating the beam are provided by the PPFX reweighting framework~\cite{ref:AliagaThesis}.  Additional transport effects are included by considering the uncertainties in the positioning and modelling of all beamline components including target, focusing horns and decay volume.  The high-energy antineutrinos in the spectrum are produced from kaons, rather than pions, and these hadrons have a 30\% uncertainty on their simulated production.  The magnitude of this is motivated by the observed discrepancy with the PPFX prediction in previous MINOS+ studies~\cite{ref:MINOS2019}.

The fraction of energy carried by neutrons that results in visible signals is uncertain.  This is quantified by considering the difference in reconstructed energy between data and simulation for quasielastic-like CC events with one $\mu^{+}$ and a single additional reconstructed object in the final state.  This second particle is selected using criteria consistent with having a neutron parent based on energy, displacement from the vertex and angle.  The observed deficit in reconstructed energy for these neutrons is used to estimate the missing energy scale.  An uncertainty on the NC-candidate event spectrum is then evaluated using this information to determine the underestimation of the reconstructed antineutrino energy from neutrons.

\begin{figure}
    \centering
    \includegraphics[width=\linewidth]{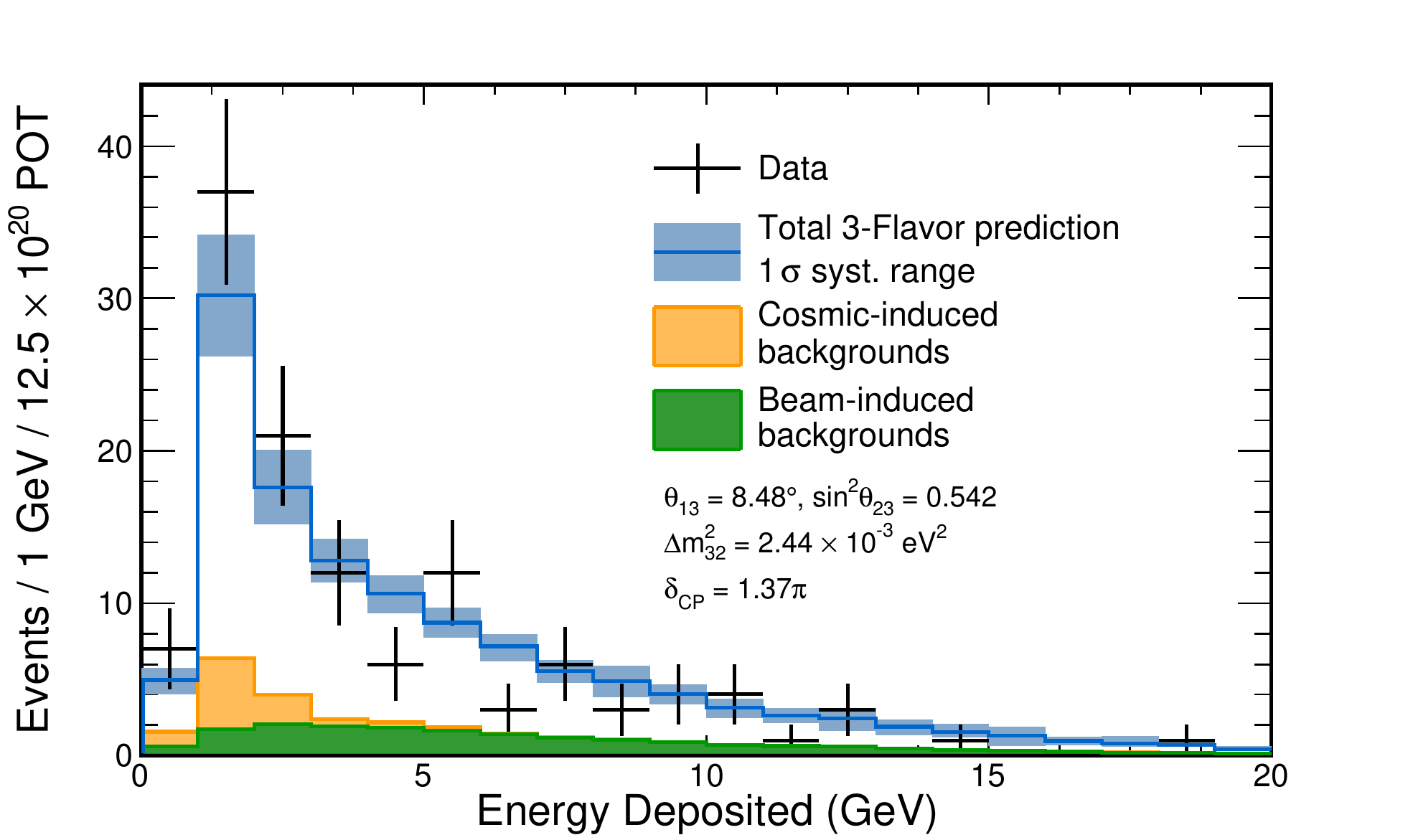}
    \caption{The reconstructed energy deposited in NC-candidate events in the FD.  The observed data spectrum is shown along with the total prediction assuming standard 3-flavor oscillations.  The contribution from the main backgrounds is also shown.}
    \label{fig:FDSpectrum}
\end{figure}

Each systematic uncertainty is applied to the simulation and used to produce oscillated FD spectra using the data-driven approach previously described.  The resulting changes in the event spectrum are taken as the residual energy shifts and included for both signal and background events as additional parameters in the fit.  The total uncertainty on simulated FD events following this procedure is shown in Fig.~\ref{fig:FDSpectrum}.

\begin{table}[th]
  \begin{centering}
  \caption[Systematic Uncertainty Summary]{The effect of the systematic uncertainties on the NC and CC expected event rates.  The difference is measured as the total change between the shifted event spectrum following the effects of a systematic with respect to the nominal spectrum.}
  \begin{tabularx}{\linewidth}{X >{\centering\arraybackslash}X >{\centering\arraybackslash}X}
    \hline\hline
                            & NC signal       & CC background   \\
    Uncertainty             & difference (\%) & difference (\%) \\
    \hline
    Calibration             & 13.8 & 9.1 \\
    Detector Response       & 4.9 & 3.8 \\
    $\nu$ Interactions      & 4.1 & 10.8 \\
    Beam                    & 1.7 & 1.3 \\
    Neutron Response        & 0.5 & 0.2 \\
    Tau Cross-Section       & - & 7.6 \\
    \hline\hline
    Total                   & 15.3 & 16.5 \\
    \hline\hline
  \end{tabularx}
  \label{tab:SystSummary}
  \end{centering}
\end{table}

The event selection and treatment of systematic uncertainties were finalized based on simulation before comparing with the FD data.  Upon analyzing the data, 121 NC-candidate events were observed, consistent with expectations from 3-flavor oscillations of $122\pm11\textnormal{(stat.)}\pm15\textnormal{(syst.)}$.  The composition of events selected from simulation is detailed in Table~\ref{tab:ExtrapNumbers}.  World-average 3-flavor oscillation parameters were used~\cite{ref:pdg}, assuming the most conservative case of normal mass ordering and upper $\theta_{23}$ octant.  The observed spectrum in Fig.~\ref{fig:FDSpectrum} shows good agreement between data and prediction across the range of energies.


\begin{table}[tb]
  \caption[Far Detector NC-Candidate Events]{Predicted number of events selected as NC candidates at the Far Detector using the data-driven procedure.  The systematic uncertainty on each of the components is shown.}
    \begin{tabularx}{0.7\linewidth}{X S[separate-uncertainty]}
    \hline\hline
    NC signal                  & 95.5(146)  \\
    $\nu_{\mu}$~CC background  & 12.2(20)   \\
    $\nu_{e}$~CC background    & 3.6(06)    \\
    $\nu_{\tau}$~CC background & 2.2(04)    \\
    Cosmic background          & 8.7(04)    \\
    \hline
    Total                      & 122.2(148) \\
    \hline\hline
  \end{tabularx}
  \label{tab:ExtrapNumbers}
\end{table}


To provide a model-independent metric of agreement with an expectation assuming oscillations between only the three active antineutrino species, the ratio $R_{\textnormal{NC}} \equiv \frac{N-\sum{P_{\textnormal{bkg}}}}{P_{\textnormal{{NC}}}}$~\cite{ref:MINOS2008_PRL}
is used to compare the number of data events, $N$, with the predicted number of signal and background events, $P_{\textnormal{NC}}$ and $P_{\textnormal{bkg}}$ respectively.  This is equal to unity in the case of no NC-disappearance.  The obtained value of $0.99\pm0.12\textnormal{(stat.)}^{+0.14}_{-0.16}\textnormal{(syst.)}$ is consistent with no oscillations involving a sterile antineutrino.

\begin{figure}[tb]
  \centering
  \includegraphics[width=\linewidth]{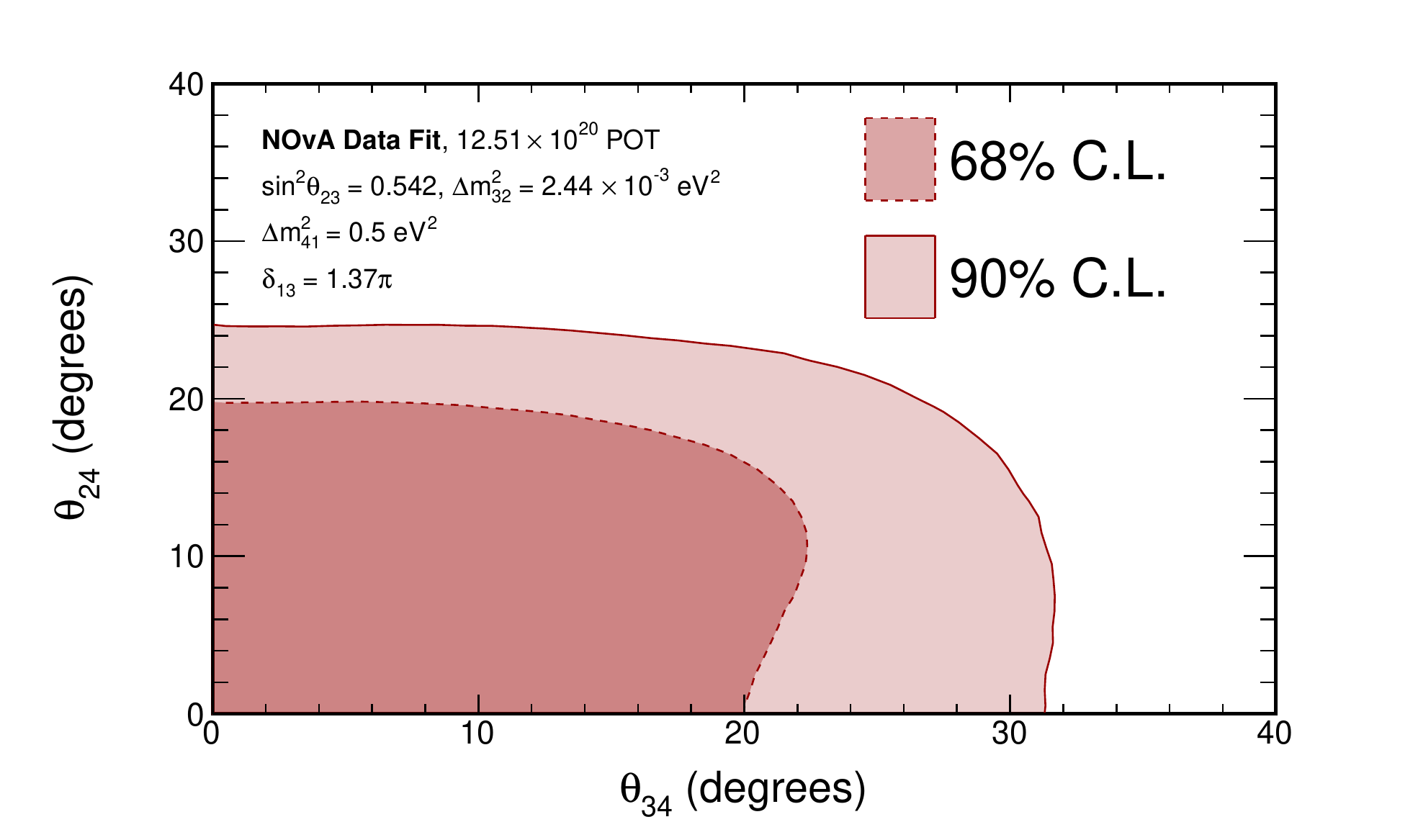}
  \includegraphics[width=\linewidth]{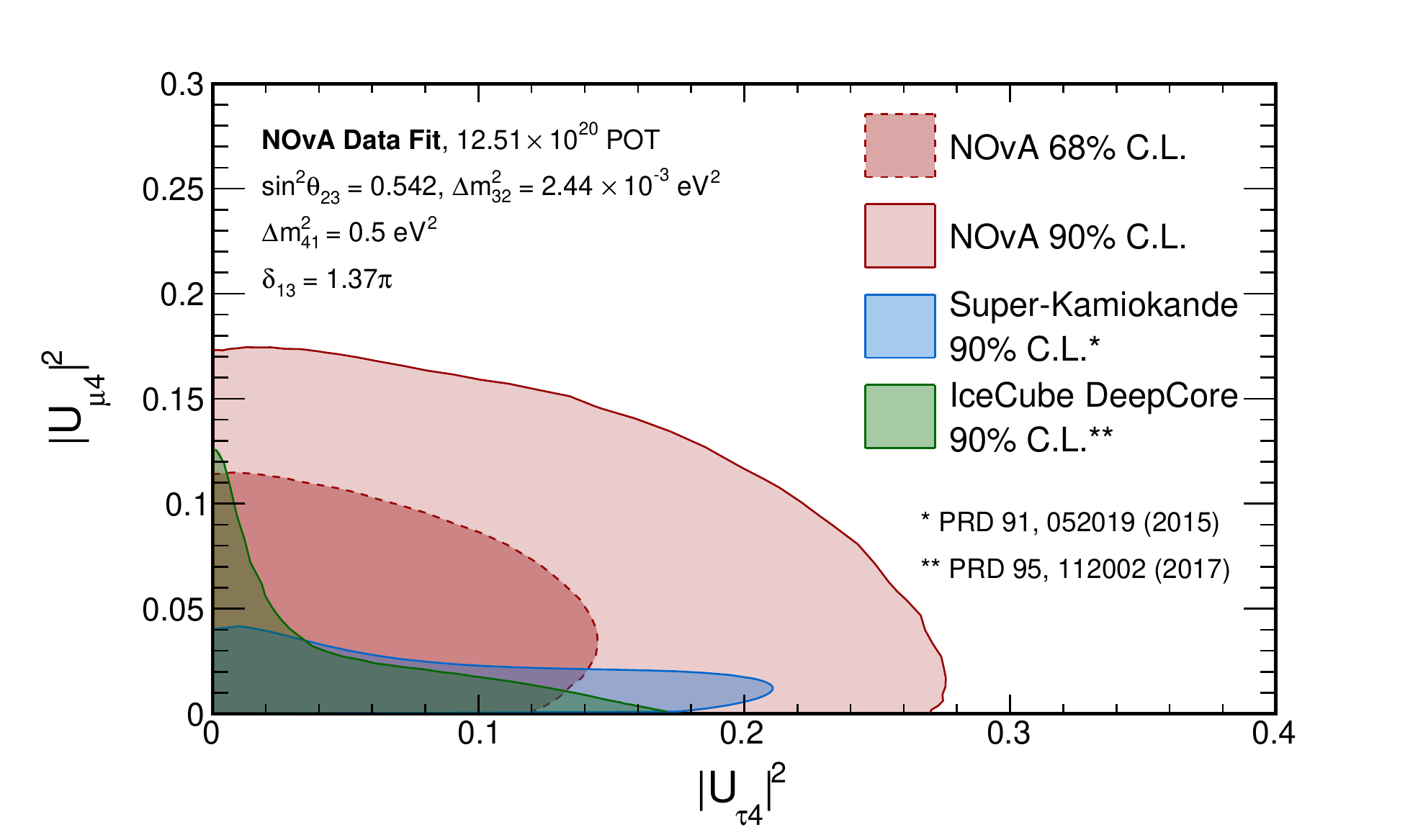}
  \caption{Allowed regions in parameter space defined by the mixing elements describing neutrino oscillations in a 3+1 model.  Top: The 68\% and 90\% allowed values of the mixing angles $\theta_{24}$ and $\theta_{34}$.  Bottom: The matrix element space, with limits from comparable measurements from Super-Kamiokande~\cite{ref:SuperKSterile2015} and IceCube DeepCore~\cite{ref:DeepCore2017}.  The limits from Super-Kamiokande and IceCube are set using atmospheric neutrinos, a sample which is majority neutrinos but with a non-negligible antineutrino component.  The NOvA limits are the first set from studying an antineutrino-dominated sample.}
  \label{fig:2DSurface}
\end{figure}

A fit considering rate and shape is performed on the spectrum in Fig.~\ref{fig:FDSpectrum} to limit the allowed values of the parameters governing sterile oscillations in the 3+1 framework.  The metric
\begin{equation}
    \chi^2 = \sum_i{2\left(P_i-N_i+N_i\ln{\frac{N_i}{P_i}}\right)} + \sum_j{\left(\frac{\Delta U_{j}}{\sigma_j}\right)^2}
\end{equation}
is minimized, varying the number of events, $P_i$, predicted in each FD bin, $i$, and constraining the agreement with data, $N_i$, using Gaussian-distributed penalty terms, $U_{j}$, for each systematic uncertainty $j$, each with width $\sigma_j$.  The mixing angle $\theta_{14}$ and the corresponding phase $\delta_{14}$ are set to zero, consistent with results from solar and reactor experiments~\cite{ref:DayaBaySterile2015} and unitarity considerations~\cite{ref:Parke2016}, and the $\delta_{24}$ CP-violating phase was included as a nuisance parameter in the fit, allowed to float freely without penalty.  The $\delta_{24}$ parameter weakens the limit for large values of both $\theta_{24}$ and $\theta_{34}$ but does not significantly impact the individual limits.  The fit is performed at $\Delta m_{41}^2 = 0.5$\,eV$^2$ but the result is valid for the full range $0.05\,\textnormal{eV}^2 \le \Delta m_{41}^2 \le 0.5\,\textnormal{eV}^2$.  Confidence limits are computed using the unified approach of Feldman-Cousins~\cite{ref:FeldmanCousins1998} and are shown in Fig.~\ref{fig:2DSurface}. 

Profiling over each sterile mixing angle in turn provides limits of $\theta_{24}<25^{\circ}$ and $\theta_{34}<32^{\circ}$ at 90\%\,C.L.  Table~\ref{tab:1DLimits} shows these constraints, together with the inferred matrix element values, alongside results from other experiments.  This result is compatible with world limits from analyses reporting no evidence of sterile neutrino oscillations, shown in Fig.~\ref{fig:2DSurface}, and is the first from an analysis of antineutrinos over a long-baseline.

\begin{table}[b]
  \begin{centering}
  \caption[Limits]{Upper 90\% C.L. limits on the mixing angles $\theta_{24}$ and $\theta_{34}$, with the associated values of the matrix elements, for this result (NOvA~$\bar{\nu}$) and comparable measurements from NOvA~$\nu$~\cite{ref:NOvASterile2017}, MINOS+~\cite{ref:MINOS2019}, T2K~\cite{ref:T2KSterile2019}, Super-Kamiokande~\cite{ref:SuperKSterile2015}, IceCube~\cite{ref:IceCube2016} and IceCube-DeepCore~\cite{ref:DeepCore2017}.  The limits are shown for $\Delta m_{41}^2 = 0.5$\,eV$^2$ for all experiments, except IceCube-DeepCore and T2K, where the results are reported for $\Delta m_{41}^2=1.0$\,eV$^2$.}
  \begin{align*}
  \begin{tabular}{p{1.3in} p{0.4in} p{0.4in} p{0.4in} p{0.4in}}
  \hline\hline
                        & $\theta_{24}$  & $\theta_{34}$  & $|U_{\mu4}|^{2}$ & $|U_{\tau4}|^{2}$ \\ \hline
NOvA ($\bar{\nu}$)      & $25^{\circ}$ & $32^{\circ}$ & 0.175            & 0.276             \\
NOvA ($\nu$)            & $20.8^{\circ}$ & $31.2^{\circ}$ & 0.126            & 0.268             \\
MINOS/MINOS+            & $ 4.4^{\circ}$ & $23.6^{\circ}$ & 0.006            & 0.160             \\
T2K                     & $18.4^{\circ}$ & $45.0^{\circ}$ & 0.1              & 0.5               \\
Super-Kamiokande & $11.7^{\circ}$ & $25.1^{\circ}$ & 0.041            & 0.180             \\
IceCube                 & $ 4.1^{\circ}$ & -              & 0.005            & -                 \\
IceCube-DeepCore        & $19.4^{\circ}$ & $22.8^{\circ}$ & 0.11             & 0.150             \\
  \hline\hline
  \end{tabular}
  \end{align*}
  \label{tab:1DLimits}
  \end{centering}
\end{table}

To conclude, 121 neutral-current candidates are observed at the Far Detector compared with $122\pm11\mbox{(stat.)}\pm15\mbox{(syst.)}$ events predicted assuming mixing only occurs between active species.  This is the first accelerator-based search for oscillations into sterile neutrinos over long-baselines using an antineutrino-dominated beam, and no evidence of a depletion of the active flavors is observed.  The measurement reported in this Letter enhances the understanding of antineutrino oscillations in the context of sterile neutrino anomalies.  Looking forward, NOvA plans to increase its antineutrino dataset by a factor of around 2.5, complementing a comparable neutrino sample, which will facilitate improved sterile neutrino searches and enable more precise measurements of the governing parameters.


\subsection*{ACKNOWLEDGEMENTS}

This document was prepared by the NOvA collaboration using the resources of the Fermi National Accelerator Laboratory (Fermilab), a U.S. Department of Energy, Office of Science, HEP User Facility. Fermilab is managed by Fermi Research Alliance, LLC (FRA), acting under Contract No. DE-AC02-07CH11359. This work was supported by the U.S. Department of Energy; the U.S. National Science Foundation; the Department of Science and Technology, India; the European Research Council; the MSMT CR, GA UK, Czech Republic; the RAS, MSHE, and RFBR, Russia; CNPq and FAPEG, Brazil; UKRI, STFC and the Royal Society, United Kingdom; and the state and University of Minnesota.  We are grateful for the contributions of the staffs of the University of Minnesota at the Ash River Laboratory, and of Fermilab.

\bibliography{nova_nc}

\end{document}

%% file: novancsterile20.tex
\newcommand{\ANL}{Argonne National Laboratory, Argonne, Illinois 60439, 
USA}
\newcommand{\ICS}{Institute of Computer Science, The Czech 
Academy of Sciences, 
182 07 Prague, Czech Republic}
\newcommand{\IOP}{Institute of Physics, The Czech 
Academy of Sciences, 
182 21 Prague, Czech Republic}
\newcommand{\Atlantico}{Universidad del Atlantico,
Carrera 30 No. 8-49, Puerto Colombia, Atlantico, Colombia}
\newcommand{\BHU}{Department of Physics, Institute of Science, Banaras 
Hindu University, Varanasi, 221 005, India}
\newcommand{\UCLA}{Physics and Astronomy Department, UCLA, Box 951547, Los 
Angeles, California 90095-1547, USA}
\newcommand{\Caltech}{California Institute of 
Technology, Pasadena, California 91125, USA}
\newcommand{\Cochin}{Department of Physics, Cochin University
of Science and Technology, Kochi 682 022, India}
\newcommand{\Charles}
{Charles University, Faculty of Mathematics and Physics,
 Institute of Particle and Nuclear Physics, Prague, Czech Republic}
\newcommand{\Cincinnati}{Department of Physics, University of Cincinnati, 
Cincinnati, Ohio 45221, USA}
\newcommand{\CSU}{Department of Physics, Colorado 
State University, Fort Collins, CO 80523-1875, USA}
\newcommand{\CTU}{Czech Technical University in Prague,
Brehova 7, 115 19 Prague 1, Czech Republic}
\newcommand{\Dallas}{Physics Department, University of Texas at Dallas,
800 W. Campbell Rd. Richardson, Texas 75083-0688, USA}
\newcommand{\DallasU}{University of Dallas, 1845 E 
Northgate Drive, Irving, Texas 75062 USA}
\newcommand{\Delhi}{Department of Physics and Astrophysics, University of 
Delhi, Delhi 110007, India}
\newcommand{\JINR}{Joint Institute for Nuclear Research,  
Dubna, Moscow region 141980, Russia}
\newcommand{\Erciyes}{
Department of Physics, Erciyes University, Kayseri 38030, Turkey}
\newcommand{\FNAL}{Fermi National Accelerator Laboratory, Batavia, 
Illinois 60510, USA}
\newcommand{\UFG}{Instituto de F\'{i}sica, Universidade Federal de 
Goi\'{a}s, Goi\^{a}nia, Goi\'{a}s, 74690-900, Brazil}
\newcommand{\Guwahati}{Department of Physics, IIT Guwahati, Guwahati, 781 
039, India}
\newcommand{\Harvard}{Department of Physics, Harvard University, 
Cambridge, Massachusetts 02138, USA}
\newcommand{\Houston}{Department of Physics, 
University of Houston, Houston, Texas 77204, USA}
\newcommand{\IHyderabad}{Department of Physics, IIT Hyderabad, Hyderabad, 
502 205, India}
\newcommand{\Hyderabad}{School of Physics, University of Hyderabad, 
Hyderabad, 500 046, India}
\newcommand{\IIT}{Illinois Institute of Technology,
Chicago IL 60616, USA}
\newcommand{\Indiana}{Indiana University, Bloomington, Indiana 47405, 
USA}
\newcommand{\INR}{Institute for Nuclear Research of Russia, Academy of 
Sciences 7a, 60th October Anniversary prospect, Moscow 117312, Russia}
\newcommand{\Iowa}{Department of Physics and Astronomy, Iowa State 
University, Ames, Iowa 50011, USA}
\newcommand{\Irvine}{Department of Physics and Astronomy, 
University of California at Irvine, Irvine, California 92697, USA}
\newcommand{\Jammu}{Department of Physics and Electronics, University of 
Jammu, Jammu Tawi, 180 006, Jammu and Kashmir, India}
\newcommand{\Lebedev}{Nuclear Physics and Astrophysics Division, Lebedev 
Physical 
Institute, Leninsky Prospect 53, 119991 Moscow, Russia}
\newcommand{\Magdalena}{Universidad del Magdalena, Carrera 32 No 22 -– 08 
Santa Marta, Colombia}
\newcommand{\MSU}{Department of Physics and Astronomy, Michigan State 
University, East Lansing, Michigan 48824, USA}
\newcommand{\Crookston}{Math, Science and Technology Department, University 
of Minnesota Crookston, Crookston, Minnesota 56716, USA}
\newcommand{\Duluth}{Department of Physics and Astronomy, 
University of Minnesota Duluth, Duluth, Minnesota 55812, USA}
\newcommand{\Minnesota}{School of Physics and Astronomy, University of 
Minnesota Twin Cities, Minneapolis, Minnesota 55455, USA}
\newcommand{\Mississippi}{University of Mississippi, University, Mississippi 38677, USA}
\newcommand{\NISER}{National Institute of Science Education and Research,
Khurda, 752050, Odisha, India}
\newcommand{\Oxford}{Subdepartment of Particle Physics, 
University of Oxford, Oxford OX1 3RH, United Kingdom}
\newcommand{\Panjab}{Department of Physics, Panjab University, 
Chandigarh, 160 014, India}
\newcommand{\Pitt}{Department of Physics, 
University of Pittsburgh, Pittsburgh, Pennsylvania 15260, USA}
\newcommand{\QMU}{School of Physics and Astronomy,
 Queen Mary University of London,
London E1 4NS, United Kingdom}
\newcommand{\RAL}{Rutherford Appleton Laboratory, Science 
and 
Technology Facilities Council, Didcot, OX11 0QX, United Kingdom}
\newcommand{\SAlabama}{Department of Physics, University of 
South Alabama, Mobile, Alabama 36688, USA} 
\newcommand{\Carolina}{Department of Physics and Astronomy, University of 
South Carolina, Columbia, South Carolina 29208, USA}
\newcommand{\SDakota}{South Dakota School of Mines and Technology, Rapid 
City, South Dakota 57701, USA}
\newcommand{\SMU}{Department of Physics, Southern Methodist University, 
Dallas, Texas 75275, USA}
\newcommand{\Stanford}{Department of Physics, Stanford University, 
Stanford, California 94305, USA}
\newcommand{\Sussex}{Department of Physics and Astronomy, University of 
Sussex, Falmer, Brighton BN1 9QH, United Kingdom}
\newcommand{\Syracuse}{Department of Physics, Syracuse University,
Syracuse NY 13210, USA}
\newcommand{\Tennessee}{Department of Physics and Astronomy, 
University of Tennessee, Knoxville, Tennessee 37996, USA}
\newcommand{\Texas}{Department of Physics, University of Texas at Austin, 
Austin, Texas 78712, USA}
\newcommand{\Tufts}{Department of Physics and Astronomy, Tufts University, Medford, 
Massachusetts 02155, USA}
\newcommand{\UCL}{Physics and Astronomy Department, University College 
London, 
Gower Street, London WC1E 6BT, United Kingdom}
\newcommand{\Virginia}{Department of Physics, University of Virginia, 
Charlottesville, Virginia 22904, USA}
\newcommand{\WSU}{Department of Mathematics, Statistics, and Physics,
 Wichita State University, 
Wichita, Kansas 67206, USA}
\newcommand{\WandM}{Department of Physics, William \& Mary, 
Williamsburg, Virginia 23187, USA}
\newcommand{\Wisconsin}{Department of Physics, University of 
Wisconsin-Madison, Madison, Wisconsin 53706, USA}
\newcommand{\deceased}{Deceased.}
\affiliation{\ANL}
\affiliation{\Atlantico}
\affiliation{\BHU}
\affiliation{\Caltech}
\affiliation{\Charles}
\affiliation{\Cincinnati}
\affiliation{\Cochin}
\affiliation{\CSU}
\affiliation{\CTU}
\affiliation{\Delhi}
\affiliation{\Erciyes}
\affiliation{\FNAL}
\affiliation{\UFG}
\affiliation{\Guwahati}
\affiliation{\Harvard}
\affiliation{\Houston}
\affiliation{\Hyderabad}
\affiliation{\IHyderabad}
\affiliation{\IIT}
\affiliation{\Indiana}
\affiliation{\ICS}
\affiliation{\INR}
\affiliation{\IOP}
\affiliation{\Iowa}
\affiliation{\Irvine}
\affiliation{\JINR}
\affiliation{\Lebedev}
\affiliation{\Magdalena}
\affiliation{\MSU}
\affiliation{\Duluth}
\affiliation{\Minnesota}
\affiliation{\Mississippi}
\affiliation{\NISER}
\affiliation{\Panjab}
\affiliation{\Pitt}
\affiliation{\QMU}
\affiliation{\SAlabama}
\affiliation{\Carolina}
\affiliation{\SMU}
\affiliation{\Stanford}
\affiliation{\Sussex}
\affiliation{\Syracuse}
\affiliation{\Texas}
\affiliation{\Tufts}
\affiliation{\UCL}
\affiliation{\Virginia}
\affiliation{\WSU}
\affiliation{\WandM}
\affiliation{\Wisconsin}

\author{M.~A.~Acero}
\affiliation{\Atlantico}

\author{P.~Adamson}
\affiliation{\FNAL}



\author{L.~Aliaga}
\affiliation{\FNAL}






\author{N.~Anfimov}
\affiliation{\JINR}


\author{A.~Antoshkin}
\affiliation{\JINR}


\author{E.~Arrieta-Diaz}
\affiliation{\Magdalena}

\author{L.~Asquith}
\affiliation{\Sussex}


\author{A.~Aurisano}
\affiliation{\Cincinnati}


\author{A.~Back}
\affiliation{\Indiana}
\affiliation{\Iowa}

\author{C.~Backhouse}
\affiliation{\UCL}

% ...

\collaboration{The NOvA Collaboration}
\noaffiliation
